\documentclass[12pt]{iopart}

\usepackage{iopams}
\usepackage{graphicx}
\usepackage{subfigure}
\usepackage{color}
\usepackage{latexsym}
\usepackage{array}
\usepackage{bbding}

\begin{document}

\title{Discrete time-crystalline order in  Bose-Hubbard model with dissipation}
\author{C. M. Dai, Z. C. Gu,  and X. X. Yi}
\address{Center for Quantum Sciences and School of Physics,
Northeast Normal University, Changchun 130024, China}
\ead{yixx@nenu.edu.cn}

\begin{indented}
	\item[]January 2020
\end{indented}

\begin{abstract}
Periodically driven quantum systems manifest various non-equilibrium features which are absent
at equilibrium. For example, discrete time-translation symmetry can be broken in periodically driven quantum
systems leading to an exotic phase of matter, called discrete time crystal(DTC). For open quantum systems, previous studies showed that  DTC can be found only when there exists  a meta-stable state in the undriven system. However, by investigating the simplest  Bose-Hubbard model with dissipation and time periodically tunneling, we find in this paper that a $2T$ DTC can appear even when the meta-stable state is absent in the undriven system. This observation   extends the understanding of DTC and shed more light on the physics behind the DTC. Besides, by the detailed analysis of simplest two-sites model, we  show further  that the two-sites model can be used as basic building blocks to construct large rings in which   a $nT$ DTC might appear. These results might find applications into engineering exotic phases in driven open quantum systems.
\end{abstract}

%
%
%
%
%


\section{Introduction}
As an analog of the spatial symmetry breaking  that leads to the formation of crystals, in 2012 Frank Wilczek proposed that time translation symmetry can be spontaneously broken in a similar  way \cite{wilczek2012}, leading to a new phase called  time crystal. The proposal of such a time crystal for time-independent Hamiltonians inspires  a lot of discussions \cite{li2012}, and those studies found  that such structures cannot exist in the ground state or any thermal equilibrium states of a quantum  system \cite{watanabe2015}, because  quantum equilibrium states are time-independent, regardless of that  the spontaneous breaking of continues time translation symmetry can occur for an excited eigenstate \cite{syrwid2017}. By comparison,    periodically driven Floquet systems posses discrete time translation symmetry, and this symmetry can be further broken into super-lattice structures where physical observables exhibit a period larger than that of the drive \cite{else2016}. The early proposal for the realization of discrete time crystal (DTC) require strong disorder to stabilize the DTC phase \cite{else2016,yao2017}. Recently, the authors of Ref.\cite{huang2018}  pointed out that the DTC can generally exist in systems without disorder. The generalization of the concept of DTC to discrete time quasicrystals is also presented  recently\cite{giergiel2019}.

In practice,   an ideally isolated system actually does not exist, and the  coupling of system to external environment may destroy the rigid time crystal behavior after a long time. This has been studied  theoretically in  Ref.\cite{lazarides2017} and experimentally in Ref.\cite{zhang2017,choi2017}. One then may wonder whether a robust time crystal order can be found in open system. The extension of DTC to open quantum system with drive and dissipation attracts widespread research interest recently. Concrete models \cite{gong2018,gambetta2019} shown that the DTC order can exist in  open system with appropriately  engineered drive and dissipation. This prediction was confirmed by the authors of Ref.\cite{gong2018} who shown that the modified Dicke model with the help of  sufficiently strong atom-photon coupling and photon loss can exhibit time crystalline structure, and the same prediction was reported  in Ref.\cite{gambetta2019} that a dissipative Rydberg model is also possible to possess  time crystal order.

For DTCs in open systems, there exist some physical observable $\hat{O}$ with expectation value $O(t)$, and $O(t)=O(t+nT)$ shows sub-harmonic response to the driving field of period $T$. Here $n$ is an integer and $n\geq2$. The Dicke model exhibits a zero-temperature phase transition from a normal to a super-radiant phase \cite{gong2018,baumann2011} when the light-matter interaction increase. The parity symmetry of the Dicke model is spontaneously broken in the super-radiant phase \cite{gong2018,baumann2011}. Ref.\cite{gong2018} utilizes this feature and drives the system in the super-radiant phase to entail sub-harmonic dynamical responses. The sub-harmonic response of Ref.\cite{gambetta2019} originates from the coexistence of two phases connected by first-order phase transition in  dissipative Rydberg gas. Though the physical realizations of these two proposals are different, both of them feature period-doubling $n=2$ and the appearance of DTC in these proposals relies on the existence of meta-stable state of the undriven system.

It is well known that  Bose-Einstein condensates (BECs) in a double-well can exhibit  macroscopic quantum self-trapping due to the inter-atomic interaction (nonlinear self-interaction in the mean-field level)\cite{smerzi1997}. Within a mean-field framework, the effect of decoherence on the dynamics of BECs in a double-well potential was  studied in Ref.\cite{wang2007}. It is  found that  the self-trapping can be either enhanced or spoiled by dissipation  depending  on the specific form of condensate-environment coupling\cite{wang2007}. This stimulate us to study how  BECs in a double-well behave under periodic drive while they are subject to decoherence, and particularly whether  the time crystal order can be found in such a system.

In this work, we introduce a Bose-Hubbard model with quasi-local dissipation and time periodic modulated tunneling. This model  describes  bosonic atoms tunneling in an optical lattice immersing in a large Bose-Einstein condensate of atoms \cite{diehl2008}. This model  can be realized in optical lattice with current technology, for details, we refer to \cite{bloch2008,diehl2008}. When the time periodic modulation is not switched on and the interaction is strong enough, our model can exhibit self-trapping, and the moderate quasi-local dissipation can help stabilize this system.

When the system is in the self-trapping regime, we find in this paper  that the system can exhibit sub-harmonic responses to the periodic modulation, a hallmark of DTC. This corresponds to the case where a meta-stable state in the undriven system exists, it is  necessary to have DTC. We further find that even the undriven system is not in the self-trapping regime, the periodic drive can also turn the system to a DTC, this interesting  result is quite different from the earlier studies, for example in Ref.\cite{gambetta2019} the appearance of DTC requires meta-stable states in the undriven system. In addition, we show that period-$n$ ($n>2$) DTC can be realized in large rings based on  two-site systems considered here. Though the period-$n$ DTCs have already been demonstrated in the closed quantum system \cite{giergiel2018},  the period-$n$ DTC has never been discussed in   open quantum systems,  and the   implementation of such a system is also lack. The realization of a period-$n$ ($n>2$) DTC in an open system is the the other interesting   result of this work.

The remainder  of this  manuscript is organized as follows: In Sec. \ref{df}, we present a formal definition of time crystal for open systems. In Sec. \ref{md}, we introduce our model. In Sec. \ref{fa}, we perform Floquet analysis for our system, and in Sec. \ref{mfa} a mean-field analysis is  given. The conclusion and discussions are presented in Sec. \ref{cd}.

\section{Definition of discrete time crystal}\label{df}
Similar to the spatial crystal, spontaneously breaking of the discrete time translation symmetry leads to the concept of discrete time crystal \cite{yao2017,else2016,huang2018}. Taken an open quantum system as an example, the broken of time translation symmetry requires the existence of a physical observable $\hat{O}$ that acts as an order parameter
\begin{equation}
O(t)=\Tr[\hat{\rho}(t)\hat{O}],
\end{equation}
to satisfy the following constraints: (A) discrete time translation symmetry breaking, i.e.,  $O(t+T)\neq O(t)$,  while the Lindbladian $\mathcal{L}(t)$ that governs the evolution of system possesses discrete time translation symmetry $\mathcal{L}(t+T)=\mathcal{L}(t)$ with period $T$; (B) rigidity: $O(t)$ possesses  a fixed oscillation period without fine-tuned system parameters; (C) persistence: the nontrivial oscillation with the fixed period must persist for infinitely long time in the thermodynamic limit. Such a definition of discrete time crystal can be regarded as an open system generalization of the definition in closed system\cite{huang2018}.

\section{Model}\label{md}
We now introduce a Bose-Hubbard model with dissipation and will show that it satisfies all the constraints (A)-(C) given in Sec. II. The model describes $N$ bosonic  atoms on an one-dimensional ring lattice with sites $M=2n$, here $n$ is an integer. The dynamics of this model is described by the Lindblad master equation,
\begin{equation}\label{mastereq}
\partial_t \hat{\rho}(t)=\mathcal{L}(t)[\hat{\rho}(t)]\equiv -i[\hat{H}(t),\hat{\rho}(t)]+\mathcal{D}[\hat{\rho}(t)],
\end{equation}
where Hamiltonian $\hat{H}(t)$ reads
\begin{equation}
\hat{H}(t)=\sum_l -J_l(t)(\hat{b}^{\dagger}_l \hat{b}_{l+1}+\hat{b}^{\dagger}_{l+1} \hat{b}_l)+\alpha_l \hat{n}_l+\frac{U_l}{2}\hat{n}_l(\hat{n}_l-1),
\end{equation}
and
\begin{equation}
\mathcal{D}[*]=\sum_l \gamma_l (2\hat{c}_l *\hat{c}_l^\dagger-\{\hat{c}^{\dagger}_l\hat{c}_l,*\}).
\end{equation}
The first part of Hamiltonian represents the kinetic energy of bosons tunneling between adjacent lattice sites with amplitude $J_l (t)$. The second part represents the onsite potential, and $\alpha_l$ is the onsite potential strength. The last part represents the inter-atomic interaction, and $U_l$ is the onsite interaction strength. $\hat{b}_l$ ($\hat{b}^{\dagger}_l$) are boson annihilation (creation) operators on site $l$, and $\hat{n}_l=\hat{b}^{\dagger}_l \hat{b}_l$ are number operators. The jump operators in $\mathcal{D}[*]$ that represent the quasi-local dissipation are given by
\begin{equation}
\hat{c}_l=(\hat{b}^{\dagger}_l+\hat{b}^{\dagger}_{l+1})(\hat{b}_l-\hat{b}_{l+1}),
\end{equation}
and $\gamma_l$ is the dissipative rate. For simplicity, we set $U_l=U$, $\alpha_{2m}=0$, $\alpha_{2m-1}=\alpha$, $J_{2m}(t)=J_o(t)$, $J_{2m-1}(t)=J_e(t)$ and $\gamma_l=\gamma$. Here $m=1,2,3,...,n$.

The tunneling amplitude $J_{o,e}(t)$ are periodic functions of period $T$, i.e.,  $J_{o,e}(t+T)=J_{o,e}(t)$. Here we define driving frequency $\omega=2\pi/T$. $U>0$ and $U<0$ represents repulsive and attractive interactions, respectively. In the following, we consider attractive interactions $U\leq0$ and tunneling amplitude $J_{o,e}(t)>0$ for any given $t$.

Note that the Lindbladian $\mathcal{L}(t)$ conserves the total particle number $\hat{N}=\sum_l \hat{n}_l$. For undriven system with constant tunneling $J_{o,e}(t)=J_{o,e}(0)$, $\alpha=0$ and non-interacting atoms, the model has a many-particle dark state $|BEC\rangle=b^{\dagger N}_{q=0}|vac \rangle/\sqrt{N!}$, corresponding to a state with macroscopic occupation of quasi-momentum $q=0$. Here, $\hat{b}_q=\sum_l \hat{b}_l e^{iql}/\sqrt{M}$ is the destruction operator for quasi-momentum $q$ in the Bloch band \cite{diehl2008}. In fact, as shown in Ref.\cite{diehl2008}, this dark state is the unique steady state of this master equation, namely, all initial states will finally evolve into $|BEC\rangle$. The implementation of this model using cold atoms in optical lattices can be found in Ref.\cite{diehl2008}.

\section{Floquet analysis}\label{fa}
\subsection{Floquet basics}
Because the evolution of our model system is governed by a period $T$ Lindbladian $\mathcal{L}(t+T)=\mathcal{L}(t)$, it is convenient to define a Floquet propagator $\mathcal{V}(T,0)$,
\begin{equation}
\mathcal{V}(T,0)=\mathcal{T}\exp[\int_0^T \mathcal{L}(\tau)d\tau]\equiv \exp(\mathcal{L}_F T).
\end{equation}
Here $\mathcal{L}_F$ plays the same role as Floquet propagator $\mathcal{V}(T,0)$, and can be treated  as effective generator of the stroboscopic time evolution $\mathcal{V}(T,0)$. In fact, according to the Floquet theorem \cite{blanes2009,dai2016}, the time evolution operator can be written as $\mathcal{V}(t,0)=\mathcal{P}(t)\exp(\mathcal{L}_F t)$ with  $\mathcal{P}(t)$ satisfying  $\mathcal{P}(t)=\mathcal{P}(t+T)$. $\mathcal{P}(t)$ represents the micro-motion in one driving period, and effective generator $\mathcal{L}_F$ governs  the evolution at the  time point $nT$($n=1,2,3,...$).

With the eigenmodes $\hat{m}_j$ of Floquet propagator $\mathcal{V}(T,0)$  given by the eigenvalue equation
\begin{equation}\label{fle}
\mathcal{V}(T,0)[\hat{m}_j]=\lambda_j \hat{m}_j,
\end{equation}
and non-degenerate eigenvalues $\lambda_j$, as well as  a  given initial state $\hat{\rho}(0)$ represented by
\begin{equation}
\hat{\rho}(0)=\sum_j \eta_j \hat{m}_j,
\end{equation}
the density operator at any time instance  of integer multiple of period can be given by
\begin{equation}\label{gs}
\hat{\rho}(nT)=\mathcal{V}^n(T,0)[\hat{\rho}(0)]=\sum_j \eta_j \exp[n\theta_jT]\hat{m}_j.
\end{equation}
Here $\theta_j=\ln(\lambda_j)/T$. $\theta_j$ can also be regard as the eigenvalue of effective generator $\mathcal{L}_F$. Because the propagator $\mathcal{V}(T,0)$ is completely positive and trace-preserving \cite{vega2017}, it possesses at least one eigenvalue equal to $1$ (maybe degenerate) and  the modulus of the rest eigenvalues are less than $1$ \cite{hartmann2017}. To simplify the representation, we  label the eigenvalues of $\mathcal{V}(T,0)$ by their length in the complex plane in descending order $1=|\lambda_1|$ and $|\lambda_j|\geqslant|\lambda_{j+1}|$.

Denoting the real and image parts of $\theta_j$ as $\theta_j^{Re}$ and $\theta_j^{Im}$, respectively, we have $\theta_j^{Re}=\ln(|\lambda_j|)/T$ and $\theta_j^{Im}=\arg(\lambda_j)/T$. Because $|\lambda_j|\leq1$, $\theta_j^{Re}$ is a non-positive real number. $-\theta_j^{Re}$ can be viewed as the effective relaxation rate of mode $\hat{m}_j$, and $\tau_j=-1/\theta_j^{Re}$ is the lifetime of mode $\hat{m}_j$. A zero $\theta_j^{Re}$ means $\hat{m}_j$ has an infinite lifetime, or say $\hat{m}_j$ is stable. The mode with nonzero $\theta_j^{Re}$ has finite lifetime. The time scale for the convergence of a given initial state to the stationary state is determined by the largest finite lifetime $\max_{\tau_j\neq \infty}\{\tau_j\}$ \cite{albert2014}. The imaginary part $\theta_j^{Im}$ is defined up to add integer multiple of driving frequency and characterizes the oscillation behavior of mode $\hat{m}_j$. In the following, we call  $\theta_j^{Re}$ effective relaxation rate and $\theta_j^{Im}$ Floquet quasi-frequency.

If some eigenvalue $\lambda_j$ of Floquet propagator is degenerate with  algebraic multiplicity $\beta_{\lambda_j}$ \cite{arnol1992}, and the number of linearly independent eigenmodes corresponding to eigenvalue $\lambda_j$ (i.e. geometric multiplicity \cite{arnol1992}) is less than $\beta_{\lambda_j}$. We can find linearly independent modes of form $\hat{m}_{j;k}(t)=\sum_{l=0}^{\beta_{\lambda_j}-1}\hat{m}_{j;k}^{l}t^l$, and the number of these modes equals to $\beta_{\lambda_j}$ \cite{arnol1992}. Here $k$ denotes the different modes corresponding to the same $\lambda_j$. The dynamics of these modes is governed by equation $\mathcal{V}(nT,0)[\hat{m}_{j;k}^{0}]=\exp[n\theta_jT]\hat{m}_{j;k}(nT)$ \cite{arnol1992}. Then a given initial state can be expanded by these linearly independent modes, i.e. $\hat{\rho}(0)=\sum_{j;k} \eta_{j;k} \hat{m}_{j;k}^0$. We have $\hat{\rho}(nT)=\sum_{j;k} \eta_{j;k}\exp[n\theta_jT]\hat{m}_{j;k}(nT)$. In this case, the eigenvalue $\lambda_j$ or $\theta_j$ can also give the information about the dynamics. Since $\hat{m}_{j;k}(t)$ is a polynomial of time $t$, it can not describe oscillating behavior, and the evolution over long time is dominated by the exponential fast decay $\propto\exp[-n|\theta_j^{Re}|T]$.

Since we are seeking for a time-crystal order  related to the specific dynamics
of the system, the Floquet eigenvalues $\lambda_j$ can provide useful information.
To satisfy the constraints (A) and (C) listed in Sec. II, we need at least two long-life modes, the life time of them should approach infinity in the thermodynamic limit, and at least one of these long life modes should has a nonzero Floquet quasi-frequency. If the initial states overlap with the long life modes that have nonzero Floquet quasi-frequency, we can expect to see the persistent  oscillation in (at least) one observable of the system. The rigidity constrain (B) in Sec. II requires the oscillation frequency to be fixed. Typically, according to the Floquet formalism $\mathcal{V}(t,0)=\mathcal{P}(t)\exp(\mathcal{L}_F t)$, on top of the evolution given by $\exp(\mathcal{L}_F t)$, there is a trivial oscillation with period $T$ given by $\mathcal{P}(t)$, the nonzero Floquet quasi-frequency of the long life modes should  be compatible with the driving frequency $\omega=2\pi/T$, such that the oscillation is still periodic. This requires that the nonzero Floquet quasi-frequency of the long life modes takes the form $\theta_j^{Im}=m\omega/n$ with integer $n=2,3,...$ and $m=1,2,..., n-1$. In general, the nonzero Floquet quasi-frequency can be incommensurate with the drive, this would lead to the so called time quasi-crystal \cite{autti2018}, but this is beyond  our scope of study in this work.

\subsection{Two-site model}
We first analyze our model system for some limiting cases to gain some insights into the problem. The dimension of the Hilbert space of our model is given by $D_b=(N+M-1)!/N!(M-1)!$. It involve a linear space of dimension of $D_b^2$ to solve the spectrum of $\mathcal{V}(T,0)$. For the simplest case $M=2$, we have $D_b=1+N$, and the spectral method is feasible for relatively large number of particle $N\sim 100$. As a comparison, when $M=4$ and $6$, we have $D_b=286$ and $3003$ for much few particle $N=10$. The case with $M>2$ will only be treated in the mean-field level. In the following, we analyze the two sites model for some limiting cases to gain insights into the problem, and the numerical results for the two sites model with general parameters are also presented.

For the sake of simplicity, we employ the following form of the tunneling strength
\begin{equation}\label{sw}
J(t)=\left\{
\begin{array}{cc}
J_1 & 0\leq t<\xi \\
J_2 & \xi\leq t<T \\
\end{array}
\right. .
\end{equation}
In this two steps driving setup, the Floquet propagator takes the form $\mathcal{V}(T,0)=\exp[\mathcal{L}(\xi)(T-\xi)]\exp[ \mathcal{L}(0) \xi]$. It is known that for two sites model with constant tunneling, if the inter-atomic interaction is large with respect to the tunneling,  and the condensate-environment coupling is chosen properly, the condensate will be locked in one of the sites, depending on the initial population \cite{wang2007}. This prediction  is obtained within a mean-field framework in Ref.\cite{wang2007}, where the number of atoms in the condensates is supposed to be infinity and the quantum fluctuation is neglected, though it is desirable to study the problem with a full quantum treatment.

For the case of symmetric double well $\alpha=0$ and time-independent tunneling, we find that except one stable mode $\hat{m}_1$ of $\mathcal{V}(T,0)$ with eigenvalue $\lambda_1=0$, a meta-stable mode $\hat{m}_2$ emerges when the interaction strength is larger than a critical value. The life time of $\hat{m}_2$ is finite for finite number of atoms, and it increases exponentially as function of atom number $N$. In the case of symmetric double well, the Lindbladian $\mathcal{L}(t)$ is invariant under the transformation $\mathcal{X}$,
\begin{equation}\label{xl}
\mathcal{X}\mathcal{L}(t)\mathcal{X}^{-1}=\mathcal{L}(t).
\end{equation}
Here $\mathcal{X}$ is an unitary transformation, defined by $\mathcal{X}[*]\equiv\hat{X}(*)\hat{X}^{\dagger}$ and $\hat{X}\equiv\exp[i\pi(b_1^{\dagger} b_2+b_2^{\dagger}b_1)/2]$. So the stable mode $\hat{m}_1$ and meta-stable mode  $\hat{m}_2$ are also the eigenmode of $\mathcal{X}$. The eigenvalue of $\mathcal{X}$ is just $\pm1$, because $\mathcal{X}^2=\mathbf{1}$. Then we conclude $\mathcal{X}[\hat{m}_1]=\hat{m}_1$, because it is the steady-state over very long time. The numerical results show that  $\mathcal{X}[\hat{m}_2]=-\hat{m}_2$. Both modes have Floquet quasi-frequency zero $\theta_j^{Im}=0$, then the linear superposition of these two modes can not show oscillatory behavior, but such a linear superposition has very long lifetime. We will show later that, the linear superposition of these two modes can represent the trapping of atoms in one of two sites. The lifetime of $\hat{m}_2$ is in fact the trapping time when the quantum fluctuation is considered.

Physically, the transformation $\mathcal{X}$ represents flipping the atoms in the first site to the second site and vice versa. If we initially load the atoms in one site, then perform a flip $\mathcal{X}$ and wait some time, the atoms can relax in the another sites due to self-trapping. Repeat this process, we can observe the period-doubling oscillation of atom imbalance between the two sites. Specifically, this two steps dynamics can be given by Floquet propagator of form
\begin{equation}
\mathcal{V}_f(T,0)=\exp[\mathcal{L}(\xi)(T-\xi)]\mathcal{X}.
\end{equation}
Denote the eigenvalue of $\hat{m}_2$ by $l_2$, i.e. $\mathcal{L}(\xi)[\hat{m}_2]=l_2 \hat{m}_2$, note that the image part of $l_2$ is zero. It is straightforward to show
\begin{equation}
\mathcal{V}_f(T,0)[\hat{m}_2]=-\exp[l_2(T-\xi)]\hat{m}_2,
\end{equation}
and in the limit $N\rightarrow \infty$, we have $\mathcal{V}_f(T,0)[\hat{m}_2]=-\hat{m}_2$. $\hat{m}_2$ has Floquet quasi-frequency $\theta_2^{Im}=\omega/2$ and the effective relaxation rate $\theta_2^{Re}=l_2(1-\xi/T)$ in this case. The effective relaxation rate of the third eigenmode $\hat{m}_3$ of $\mathcal{V}_f(T,0)$ is just $\theta_3^{Re}=l_3^{Re}(1-\xi/T)$ where $l_3$ is given by $\mathcal{L}(\xi)[\hat{m}_3]=l_3 \hat{m}_3$. $l_3^{Re}$ takes a nonzero value in the large $N$ limit, so the long time dynamics of system is characterized by the first two modes. To realize such Floquet propagator, we can set $J_1 \xi=\pi/2$ in Eq.(\ref{sw}). Besides, $\gamma N \xi$ should be small enough so that dissipation in the time interval $[0,\xi)$ is negligible, and $J_1\gg|UN|$ so the unitary part of evolution is given approximately by $\mathcal{X}$. It is unclear whether the meta-stable mode with Floquet quasi-frequency $\theta_2^{Im}=\omega/2$ can
still exist or not, when the slight deviation of the first driving step $\exp[ \mathcal{L}(0) \xi]$ from $\mathcal{X}$ are considered. We study this question by numerically solving Eq.(\ref{fle}) in the following subsection.

Suppose the meta-stable mode with Floquet quasi-frequency $\theta_2^{Im}=\omega/2$ can still exist when the imperfections in the driving step are considered. If the evolution of system starts from a state $\hat{\rho}(0)=\hat{m}_1+c_2 \hat{m}_2+\sum_{k>2}c_k\hat{m}_k$ and $c_2\neq0$. After time $\tau_3=-1/\theta_{3}^{Re}$ and in the large $N$ limit, the density matrix at stroboscopic moment $t=nT$ is
\begin{equation}\label{rt}
\hat{\rho}(nT)=\hat{m}_1+(-1)^{n}c_2 \hat{m}_2,
\end{equation}
that shows sub-harmonic response to the driving field. Define the atom imbalance between the two sites,
\begin{equation}\label{op}
\hat{O}=\frac{1}{2N}(\hat{b}^{\dagger}_1\hat{b}_1-\hat{b}^{\dagger}_2\hat{b}_2).
\end{equation}
The expectation value of observable $\hat{O}$ with state given by Eq.(\ref{rt}) is
\begin{equation}\label{ont}
O(nT)=\Tr[\hat{O}\hat{\rho}(nT)]=(-1)^n c_2\Tr[\hat{m}_2 \hat{O}].
\end{equation}
This shows period-doubling oscillation, a clear signature of discrete time crystal. Here, we use the relation $\Tr[\hat{O}\hat{m_1}]=0$ to obtain Eq.(\ref{ont}), that comes from $\mathcal{X}[\hat{m}_1]=\hat{m}_1$. If the two sites are asymmetric, i.e. $\alpha\neq0$, there is no symmetry constraint to $\hat{m}_1$, and generally $\Tr[\hat{O}\hat{m_1}]\neq0$, $O(nT)$ can acquire a nonzero basis.

\subsection{Numerical results for the model with two sites}
\begin{figure}
	\centering
	\includegraphics[scale=0.48]{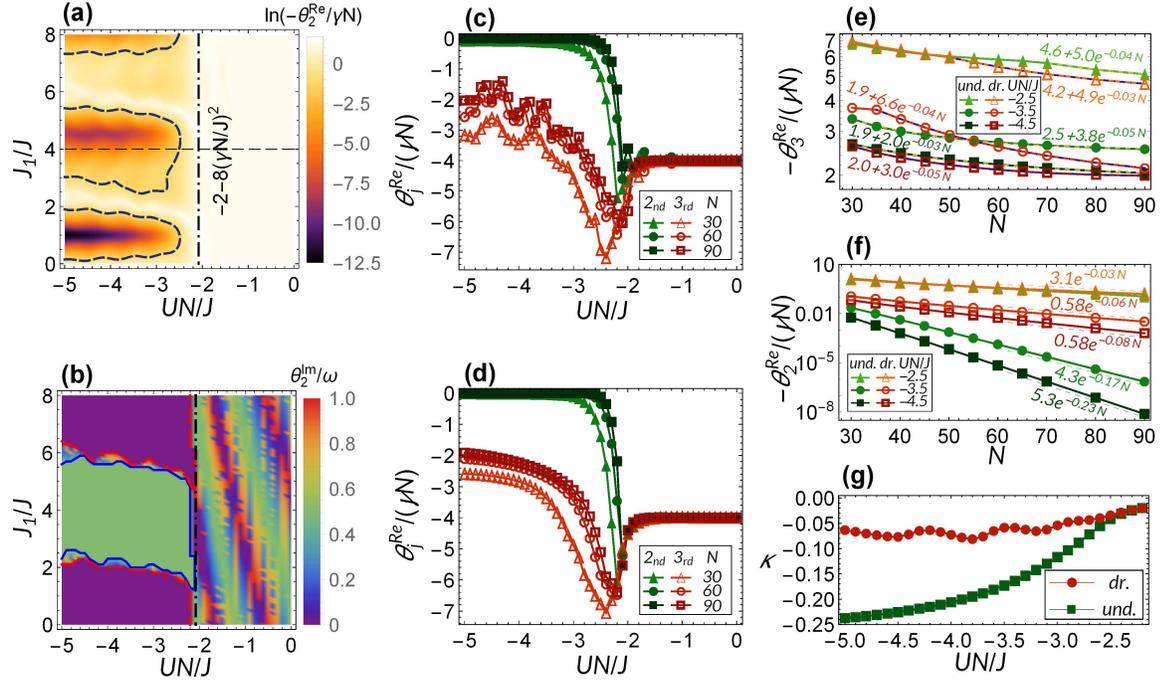}
	\caption{(a) and (b): The effective relaxation rate $\theta_2^{Re}$ and Floquet frequency $\theta_2^{Im}$ of the second eigenmode of $\mathcal{V}(T,0)$ as a function of $J_1/J$ and $UN/J$. Dotted dash line in (a) and (b) indicates the critical interaction strength $UN/J=-2-8(\gamma N/J)^2$ of parent generator $\mathcal{L}(\xi)$  obtained by the mean-field  theory. The dash lines in (a) enclose the region with $\ln[-\theta_2^{Re}/(\gamma N)]<-1$. The solid and dash lines in (b) enclose the region where $\theta_2^{Im}=\omega/2$ and $0$, respectively. The parameters used in (a) and (b) are $J_2=J=1$, $\gamma=0.1J/N$, $\alpha=0$, $T=2.5\pi/J$, $\xi=\pi/(8J)$ and the number of particle is $N=60$. (c) $\theta_2^{Re}$ and $\theta_3^{Re}$ of $\mathcal{V}(T,0)$ as a function of interaction strength $UN/J$ for three different particle numbers $N=30,60,90$, respectively. Except the tunneling strength $J_1/J=4$, the other parameters used in (c) are the same as in (a). (d) $\theta_2^{Re}$ and $\theta_3^{Re}$ of $\exp[\mathcal{L}(\xi)T]$ for three different particle numbers $N=30,60$ and $90$, the other parameters used are the same as in  (a). (e) and (f), the finite size scaling results for $\theta_3^{Re}$ and $\theta_2^{Re}$. The lines with hollow and solid markers for $\mathcal{V}(T,0)$ and $\exp[\mathcal{L}(\xi)T]$, respectively. The dash lines in (e) and (f) indicate the biased exponential fitting $a+b e^{\tilde{\kappa} N}$ and exponential fitting $c e^{\kappa N}$, respectively. Three different interaction strength $UN/J=-2.5,-3.5,-4.5$ are chosen in (e) and (f) for comparison. The other parameters used in (e) and (f) are the same as in (a). The factors $\kappa$ of the exponential fitting in (f) versus $UN/J$ are plotted in (g),  lines with dotted and square symbols are plotted for $\mathcal{V}(T,0)$ and $\exp[\mathcal{L}(\xi)T]$, respectively.}\label{scale}
\end{figure}

First, we consider two symmetric sites, i.e. $\alpha=0$. For Floquet propagator with tunneling of form Eq.(\ref{sw}), the first three eigenvalues are calculated numerically, and the results are shown in Fig.\ref{scale}. The panels (a) and (b) in Fig.\ref{scale} plot the effective relaxation rate $\theta_2^{Re}$ and Floquet quasi-frequency $\theta_2^{Im}$ of the second eigenmode of $\mathcal{V}(T,0)$ as functions of interaction strength $UN/J$ and tunneling $J_1/J$, respectively. The particle number $N$ used in the calculation of Fig.\ref{scale} (a) and (b) are $N=60$. The dotted dash line in Fig.\ref{scale} (a) and (b) indicates the mean-field critical interaction strength $U_{cr}N/J=-2-8(\gamma N/J)^2$ of parent generator $\mathcal{L}(\xi)$. The derivation of $U_{cr}$ is straightforward and can be found in section \ref{mfa}.

As shown in Fig.\ref{scale}(a), on the right hand side of the dotted dash line, the three regions with $\ln[-\theta_2^{Re}/(\gamma N)]<-1$ are surrounded  by dash lines. In these three regions, $|\theta_2^{Re}|$ is several orders of magnitude smaller than the natural relaxation rate $\gamma N$. The Floquet quasi-frequency $\theta_2^{Im}$ of the second eigenmode of $\mathcal{V}(T,0)$ in these three regions are $0$, $\omega/2$ and $0$ from up to the bottom, respectively. The appearance of meta-stable mode with Floquet quasi-frequency $\theta_2^{Im}$ in the middle region of Fig. \ref{scale} (a) and (b) indicate the time crystal phase is robust to the moderate deviation of driving parameters from the ideal case. The dash line in Fig.\ref{scale}(a) indicates $J_1\xi=\pi/2$ as a guide to the eyes. $\theta_2^{Im}=0$ in the up and bottom regions can be understood as the rigid of the meta-stable mode of $\mathcal{L}(\xi)$ under time-dependent perturbations. Since for $J_1/J\approx1$ in the bottom region, $\mathcal{L}(0)\approx\mathcal{L}(\xi)$ and $\mathcal{V}(T,0)\approx \exp[\mathcal{L}(\xi)T]$. In the case  $J_1/J=1$, this relaxation is exact, i.e. $\mathcal{V}(T,0)= \exp[\mathcal{L}(\xi)T]$. Such phenomenon is referred as time-dependent self-trapping of BECs \cite{cui2010}, and the self-trapping happens in a time-dependent state.
In our case, the time-dependent self-trapping state is the time periodic Floquet state, the atoms still mainly occupy one of the two sites, but different from the time independent case, the atoms can have micro-motion near the trapping position with period driving. As for the upper region, the parameters $J_1\approx8J$, $\mathcal{V}(\xi,0)\approx \mathbf{1}$ and $\mathcal{V}(T,0)\approx \exp[\mathcal{L}(\xi)(T-\xi)]$, the same argument also applies.

The effective relaxation rates of the second and third eigenmodes of $\mathcal{V}(T,0)$ for three different particle number $N=30,60$ and $90$ as functions of interaction strength $UN/J$ are shown in Fig.\ref{scale} (c). For comparison, the effective relaxation rates of the second and third eigenmodes of $\exp[\mathcal{L}(\xi)T]$ are also calculated and plotted in Fig.\ref{scale} (d) (the effective relaxation rates defined here are just the real part of the eigenvalues of parent generator $\mathcal{L}(\xi)$). We can see that in both Fig.\ref{scale} (c) and (d), the effective relaxation rate of the second eigenmode drops sharply to nearly zero when the interaction strength reaches a critical value $U_{cr}$. This means the second modes become meta-stable states. When the interaction strength reaches $U_{cr}$, the effective relaxation rates of the third modes also change, but are still of the same order of the natural relaxation rate $\gamma N$. The $U_{cr}$ of Fig.\ref{scale} (c) and (d) are very close for the symmetric two sites case. We find that the larger the particle number there is, the phase boundary becomes more clear. The $U$ dependent oscillation of the relaxation rate in Fig.\ref{scale} (c) can be understood from the fact $\mathcal{V}(T,0)=\exp[\mathcal{L(\xi)}(T-\xi)]\exp[\mathcal{L}(0)\xi]\approx\exp[\mathcal{L(\xi)}(T-\xi)] \mathcal{X}$. Because $\mathcal{X}\mathcal{L}(t)\mathcal{X}^{-1}=\mathcal{L}(t)$ and $\mathcal{X}^2=1$, if $\hat{m}_j$ satisfies eigenvalue equation $\exp[\mathcal{L(\xi)}T][\hat{m}_j]=\lambda_j \hat{m}_j$, we then have $\mathcal{X}[\hat{m}_j]=\pm\hat{m}_j$ and $\exp[\mathcal{L(\xi)}(T-\xi)]\mathcal{X}[\hat{m}_j]=\pm\lambda_j^{(T-\xi)/T}\hat{m}_j$. The effective relaxation rate of Floquet propagator $\mathcal{V}(T,0)$ can be approximated by $\theta_j^{Re*}=\ln(|\lambda_j^{(T-\xi)/T}|)/T=[(T-\xi)/T]\ln(|\lambda_j|)/T=\theta_j^{Re}(T-\xi)/T$. Here $\theta_j^{Re}$ denotes the relaxation rate of propagator $\exp[\mathcal{L(\xi)}T]$. Because $\mathcal{L}(0)$ is a function of $U$, but $\mathcal{X}$ is $U$ independent, with the increase of $U$, the approximation $\exp[\mathcal{L}(0)\xi]\approx\mathcal{X}$ becomes worse,  and the exact effective relaxation rate of Floquet propagator deviates from $\theta_j^{Re*}$,  leading to a $U$-dependent oscillation of the effective relaxation rate near $\theta_j^{Re*}$.

The finite size scaling results of $\theta_3^{Re}$ and $\theta_2^{Re}$ of $\mathcal{V}(T,0)$ as functions of particle number $N$ for three typical interaction strengths used in Fig.\ref{scale} (c) are plotted in Fig.\ref{scale}(e) and (f) using lines with hollow markers, respectively. The same results but for $\exp[\mathcal{L}(\xi)T]$ are also plotted in Fig.\ref{scale}(e) and (f). As shown in Fig.\ref{scale}(e), we have relaxation rate $-\theta_3^{Re}/(\gamma N)= a+b\exp[\tilde{\kappa} N]$ that approach a non-zero value $a$ exponentially. The value of $a$ is of order $1$. In Fig.\ref{scale}(f), we have relaxation rate $-\theta_2^{Re}/(\gamma N)\propto \exp[\kappa N]$. The factor $\kappa$ for $\mathcal{V}(T,0)$ and $\exp[\mathcal{L}(\xi)T]$ are plotted in Fig.\ref{scale}(g) using lines with dot and square symbols, respectively. For $\exp[\mathcal{L}(\xi)T]$, $\kappa$ monotonically decrease as the increasing of interaction strength. For $\mathcal{V}(T,0)$, $\kappa$ has more complicated relation with the interaction strength.

\begin{figure}
	\centering
	\includegraphics[scale=0.48]{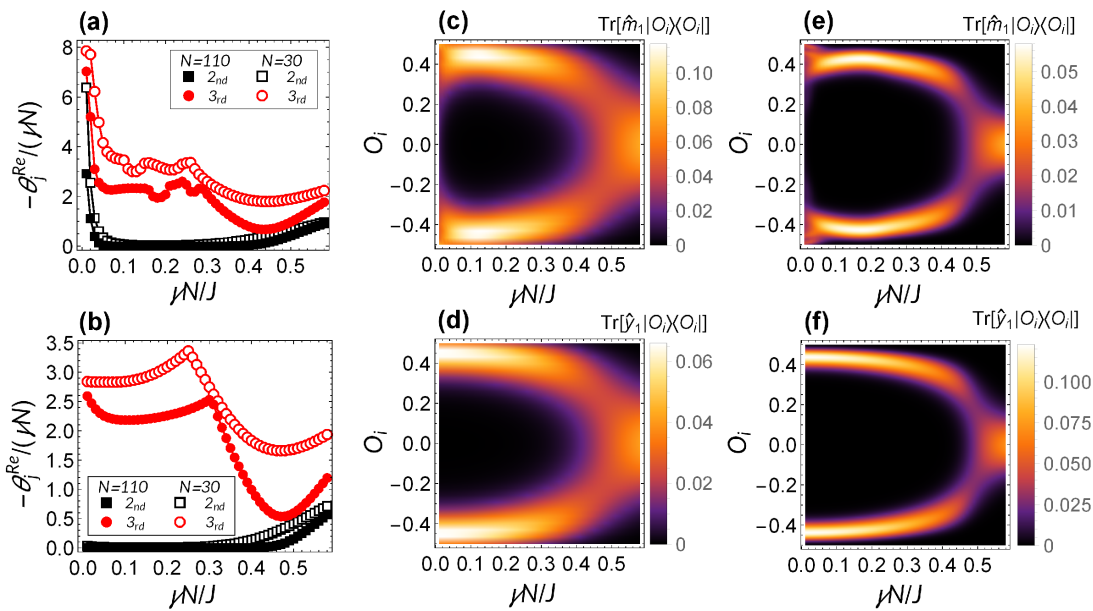}
	\caption{(a) The second and third effective relaxation rate of $\mathcal{V}(T,0)$ as a  function of dissipation strength $\gamma N/J$ for two different particle numbers $N=110$ and $30$. (b)is the  same as (a), but for propagator $\exp[\mathcal{L}(\xi)T]$. (c) and (e) show the projection of the first eigenmode of $\mathcal{V}(T,0)$ to the eigenstates of $\hat{O}$. The particle numbers used in (c) and (e) is $N=30$ and $110$, respectively. (d) and (f) are respectively the same as (c) and (e), but for the first eigenmode of $\exp[\mathcal{L}(\xi)T]$. The other parameters used are $J_2=J=1$, $J_1=4J$, $\alpha=0$, $T=2.5\pi/J$, $\xi=\pi/(8J)$ and $UN/J=-4$.}\label{gam}
\end{figure}

One may wonder what role the quasi-local dissipation plays in the appearance of time-crystal order. In Fig.\ref{gam} (a), we plot the relaxation rate of the second and third eigenmodes of $\mathcal{V}(T,0)$ as functions of quasi-local dissipation strength $\gamma N/J$. The relaxation rate of the second and third eigenmodes of propagator $\exp[\mathcal{L}(\xi)T]$ are also plotted in Fig.\ref{gam} (b) as a reference. We can see for sufficiently weak dissipation, for instance $\gamma N/J\approx0.04$, in Fig.\ref{gam}(a) there is a sharp change of $-\theta_2^{Re}/(\gamma N)$ from nearly zero to about $3$ and $6$ for $N=110$ and $30$, respectively. But the reference in Fig.\ref{gam} (b) is still nearly zero. This suggests that the meta-stable modes are more fragile when the dissipation is weak. Too strong dissipation also lifts $-\theta_2^{Re}/(\gamma N)$ from zero for both Fig.\ref{gam} (a) and (b), but the process is much slowly. As shown in Fig.\ref{gam} (a) and (b) near $\gamma N/J=0.4$, except the slowly increasing of $-\theta_2^{Re}/(\gamma N)$, there is a dip of $-\theta_3^{Re}/(\gamma N)$. The behavior of Floquet quasi-frequency versus dissipation strength is similar with Fig.\ref{scale} (b). With the increase of effective relaxation rate, the Floquet quasi-frequency is no longer fixed to a particular value.

Fig.\ref{gam} (c) and (d) show the projections of the first eigenmodes of $\mathcal{V}(T,0)$ ($\hat{m}_1$) and $\exp[\mathcal{L}(\xi)T]$ ($\hat{y}_1$) on the eigenstates $|O_i\rangle\langle O_i|$ of operator $\hat{O}$ defined by Eq.(\ref{op}), respectively. The eigenvalues $O_i$ ranges from $-0.5$ to $0.5$. The results in Fig.\ref{gam} (c) and (d) are for the case of $N=30$. Fig.\ref{gam} (e) and (f) show the same thing as (c) and (d) but for larger particle number $N=110$. Comparing the results in Fig.\ref{gam} (c)-(f) with (a)-(b), we can find the lifting of $-\theta_j^{Re}/(\gamma N)$ is accompanying with the changing of probability distribution on $|O_i\rangle\langle O_i|$. In the case of moderate dissipation, the probability distribution of $\hat{m}_1$ and $\hat{y}_1$ as a function of $O_i$ have two symmetric peaks, and the meta-stable states exit. The probability distribution on $|O_i\rangle\langle O_i|$ of linear superposition of $\hat{m}_1$ ($\hat{y}_1$)  with another meta-stable mode can have two asymmetric peaks, or in limit case only one peak that locates near $O_i=\pm0.5$. So for undriven system, the initial atoms imbalance can sustain a very long time for large particles number, and when $N\rightarrow\infty$, the initial atoms imbalance can freeze permanently. If we switch on the periodic modulation of tunneling, the initial loaded condensate in one of the two sites can jump back and forth between the two sites. When the dissipation is strong enough, the two peaks in Fig.\ref{gam} (c)-(f) merge into one peak locating at $O_i=0$ and the meta-stable mode disappears. For too weak dissipation, though the two peaks also exist for $\exp[\mathcal{L}(\xi)T]$, the periodic driving turns the two peaks to broad distributions that have rich structures, which is more clearly for larger $N$. When such structures appear, the corresponding mean-field dynamics is chaotic as show in section \ref{mfa}.

\begin{figure}
	\includegraphics[scale=0.48]{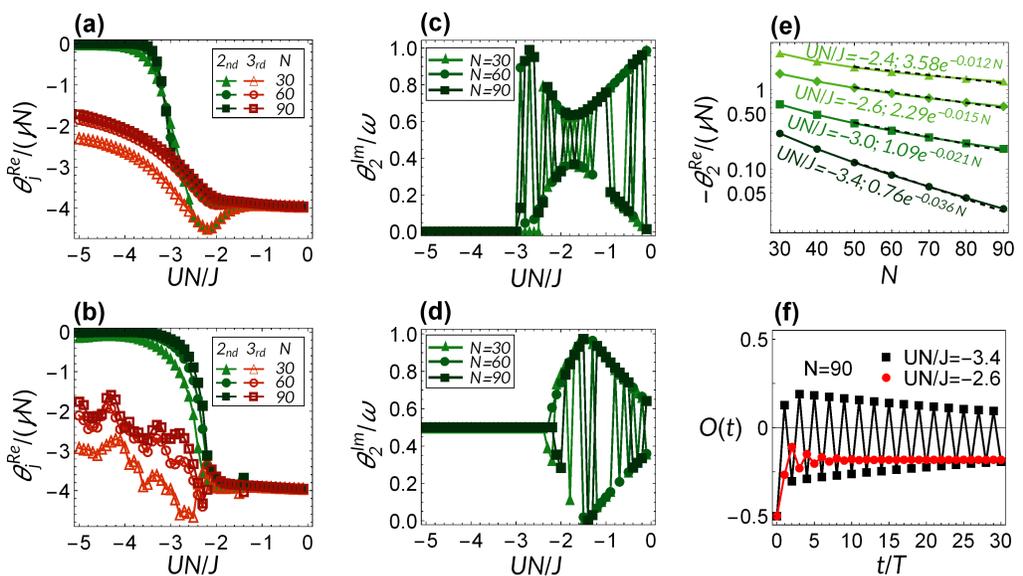}
	\caption{(a) and (b) are the effective relaxation rates of the second and third eigenmodes of propagator $\exp[\mathcal{L}(\xi)T]$ and $\mathcal{V}(T,0)$ versus  interaction strength $UN/J$. (c) and (d) are for the Floquet frequency of the second eigenmodes of $\exp[\mathcal{L}(\xi)T]$ and $\mathcal{V}(T,0)$, respectively. The parameters used in (a)-(d) are $J_2=J=1$, $J_1=4J$, $\alpha=0.3J$, $\gamma N/J=0.2$,$T=\pi/J$, $\xi=\pi/(8J)$. Three different particle numbers $N=30,60,90$ are chosen for these plots. (e) The scaling of the effective relaxation rate of the second eigenmode of $\mathcal{V}(T,0)$ versus particle number $N$ for four different interaction strength $UN/J=-3.4,-3.0,-2.6,-2.4$, the dash lines indicate an exponential fitting. The other parameters used in (e) are the same as in (b). (f) The evolution of density imbalance for two different interaction strengths $UN/J=-3.4$ (line with square symbols) and $UN/J=-2.6$ (line with dotted symbols), the other parameters used are the same as in (b), except particle number $N=90$. We assume the atoms was initially  loaded    in the second site, i.e. $\rho(0)=|0,N\rangle \langle 0,N|$.}\label{broken}
\end{figure}

In the case $\alpha\neq0$. Direct numerical results show that the meta-stable mode of $\mathcal{V}(T,0)$ with Floquet frequency $\theta_2^{Im}=\omega/2$ also exist. The effective relaxation rates of the second and third eigenmodes of  $\mathcal{V}(T,0)$ with $\alpha\neq0$ are shown in Fig.\ref{broken} (b). Fig.\ref{broken} (a) shows the same thing but for $\exp[\mathcal{L}(\xi)T]$ as a comparison. The Floquet frequency of the second eigenmode of $\exp[\mathcal{L}(\xi)T]$ and $\mathcal{V}(T,0)$ is plotted in Fig.\ref{broken} (c) and (d), respectively. We can find from Fig.\ref{broken} (a)-(d), the critical interaction strength for the appearance of meta-stable mode of Floquet propagator $\mathcal{V}(T,0)$ with quasi-frequency $\omega/2$ is smaller than the interaction strength needed for the appearance of zero frequency meta-stable mode of the corresponding $\exp[\mathcal{L}(\xi)T]$. So even the undriven system is not in the self-trapping regime, the time crystal order can emerge due to the joint effect of driving, interaction and dissipation. Fig.\ref{broken} (e) plots the scaling of the effective relaxation rate of the second eigenmode of $\mathcal{V}(T,0)$ as functions of $N$ with four different interaction strengths, that also shows exponential decay $\sim\exp[\kappa N]$. Fig.\ref{broken} (f) shows the periodic doubling oscillation of the imbalance $\hat{O}$. The initial state $\rho(0)=|0,N\rangle\langle 0,N|$ is used for the numerical simulation in Fig.\ref{broken} (f). Here we define $|k, N-k \rangle \equiv \hat{b}_1^{\dagger k} \hat{b}_2^{\dagger (N-k)}|vac\rangle/\sqrt{k!(N-k)!}$, $k =0,1,...,N$. The slightly deviation of the initial state from $|0,N\rangle\langle 0,N|$, i.e. choosing $\rho(0)=|0,N\rangle\langle 0,N|+\varepsilon \rho_p$ with small real number $\varepsilon$, can give results similar with Fig.\ref{broken} (f). Because the scaling factor $|\kappa|=0.015$ is small when the interaction strength $UN/J=-2.6$, for finite number of particle $N=90$, the corresponding relaxation time scale of the second eigenmode $\tau/T \approx 3$ is much short than the case $UN/J=-3.4$ that has a relaxation time $\tau/T\approx 53$. As shown in Fig.\ref{broken} (f), the oscillation of imbalance decays much fast for $UN/J=-2.6$ than $UN/J=-3.4$. Note that the projection of the first eigenmode of $\mathcal{V}(T,0)$ to the eigenstates of $\hat{O}$ also has double-peak structure similar with the case $\alpha=0$. The projection of the first eigenmode of $\exp[\mathcal{L}(\xi)T]$ to the eigenstates of $\hat{O}$ only has one peak and the second meta-stable mode has double-peak structure, the linear superposition of these two modes are responsible for the self-trapping feature of the undriven system.

\begin{figure}
	\includegraphics[scale=0.5]{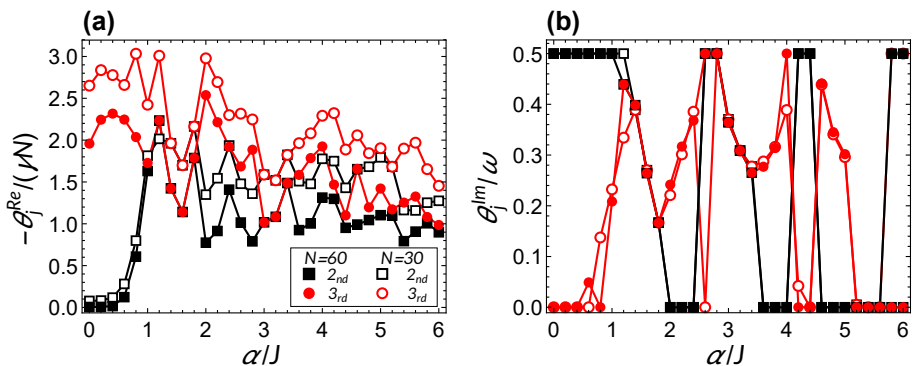}
	\caption{(a) The second and third effective relaxation rate of $\mathcal{V}(T,0)$ as a function of off-set potential $\alpha$ for two different particle numbers $N=60$ and $30$. (b) is the corresponding Floquet quasi-frequency.  The other parameters used are $J_2=J=1$, $J_1=4J$, $T=\pi/J$, $\xi=\pi/(8J)$, $\gamma N/J=0.2$ and $UN/J=-4$.}\label{alpha}
\end{figure}

We show the effective relaxation rate of $\mathcal{V}(T,0)$ as functions of off-set potential $\alpha$ in Fig. \ref{alpha}. We find that stronger $\alpha$ can lead to a sharp change in  the effective relaxation rate for the second eigenmode, and it destroys the DTC phase too. When $\alpha$ is smaller than a critical value, the Floquet quasi-frequency of the second eigenmode equals to $\omega/2$.

\section{Mean-field analysis}\label{mfa}
First, we analysis the two sites model within the mean-field framework and compare the results with previous one that consider quantum fluctuations. Then we switch to the mean-field analysis of general $2n$ sites models.

\subsection{Mean-field analysis for two sites model}
For the two-mode Bose-Hubbard model described by Eq.(\ref{mastereq}), it is convenient to introduce the spin operators defined by \cite{watanabe2012,hartmann2017}
\begin{equation}
		 \eqalign{
		\hat{S}_x=\frac{1}{2N}(\hat{b}^{\dagger}_1\hat{b}_2+\hat{b}^{\dagger}_2\hat{b}_1), \cr
		\hat{S}_y=\frac{1}{2iN}(\hat{b}^{\dagger}_1\hat{b}_2-\hat{b}^{\dagger}_2\hat{b}_1), \cr
		\hat{S}_z=\frac{1}{2N}(\hat{b}^{\dagger}_1\hat{b}_1-\hat{b}^{\dagger}_2\hat{b}_2),}
\end{equation}
These spin operators satisfy the commutation relations $[\hat{S}_\lambda,\hat{S}_\mu]=\frac{i}{N}\varepsilon_{\lambda\mu\nu}\hat{S}_\nu$ with Levi-Civita symbol $\varepsilon_{\lambda\mu\nu}$. Using these spin operators, the Hamiltonian $\hat{H}(t)$ and jump operator $\hat{c}$ can be rewritten as
\begin{equation}
	\eqalign{
		\hat{H}(t)=&-2J(t)N\hat{S}_x+\alpha N \hat{S}_z+UN^2 \hat{S}_z^2+const. ,\cr
		\hat{c}=&2N(\hat{S}_z-i\hat{S}_y),}
\end{equation}
with irrelevant constant $const.=(\alpha-U)N/2+UN^2/4$. Since the Hamiltonian $\hat{H}(t)$ and the jump operator $\hat{c}$ conserve total particle number $N$, $\hat{\textbf{S}}^2\equiv\sum_{i=x,y,z}\hat{S}_i^2=(N+2)N/4N^2$ is a constant of motion. The evolution of the expectation value of spin operators $S_i(t) \equiv \Tr[ \hat{S}_i \hat{\rho}(t)]$ can be obtained using Eq.(\ref{mastereq}) and the mean-field equation for $S_i(t)$ is acquired by the assumption $\Tr[\hat{S}_i\hat{S}_j \hat{\rho}(t)]\approx\Tr[\hat{S}_i\hat{\rho}(t)]\Tr[\hat{S}_j \hat{\rho}(t)]$ for large $N$ and neglect the term in the order of $N^{-1}$, i.e.
\begin{equation}\label{sequation}
	\eqalign{
\partial_t S_x= &-\alpha S_y-2UN S_y S_z+8\gamma N(S_z^2+S_y^2),\cr
\partial_t S_y= &2J(t) S_z+\alpha S_x +2UN S_x S_z-8\gamma N S_y S_x,\cr
\partial_t S_z= &-2J(t) S_y-8\gamma N S_z S_x,}
\end{equation}
here, we have omitted the argument $t$ in $S_i(t)$ for simplicity. The evolution of quantity $\textbf{S}^2=S_x^2+S_y^2+S_z^2$ governed by Eq.{(\ref{sequation})} is also a constant of motion and this relation reduces the three equations in Eq.{(\ref{sequation})} to two equations \cite{hartmann2017}
\begin{equation}\label{angle}
	\eqalign{
\partial_t \theta &= 2 J(t)\sin(\phi)+4\gamma N \cos(\theta)\cos(\phi),\\
\partial_t \phi &= 2J(t)\frac{\cos(\theta)}{\sin(\theta)} \cos(\phi)+\alpha+UN\cos(\theta)-4\gamma N \frac{\sin(\phi)}{\sin(\theta)},}
\end{equation}
here $\theta$ and $\phi$ are polar and azimuth angles by expressing $\textbf{S}$ in the spherical coordinate system, i.e.
\begin{equation}
\{S_x,S_y,S_z\}=\{\cos(\phi)\sin(\theta),\sin(\phi)\sin(\theta),\cos(\theta)\}/2.
\end{equation}
Without the periodically modulated tunneling, i.e. $J(t)=J$ is time-independent and $\alpha=0$, the nonlinear differential equations Eq.(\ref{sequation}) with constraint condition $S_x^2+S_y^2+S_z^2=1/4$ have six different solutions that satisfy $\partial_t S_i=0$ ($i=x,y,z$). Three of the six solutions are not real. The stability analysis of the remaining three solutions shows that two or one solutions are stable depending on the interaction strength. When $UN/J<-2-8(\gamma N/J)^2$, there are two stable solutions. For the case $UN/J>-2-8(\gamma N/J)^2$, there is one unique stable solution $\{S_x,S_y,S_z\}=\{1/2,0,0\}$.

\begin{figure*}
	\includegraphics[scale=0.46]{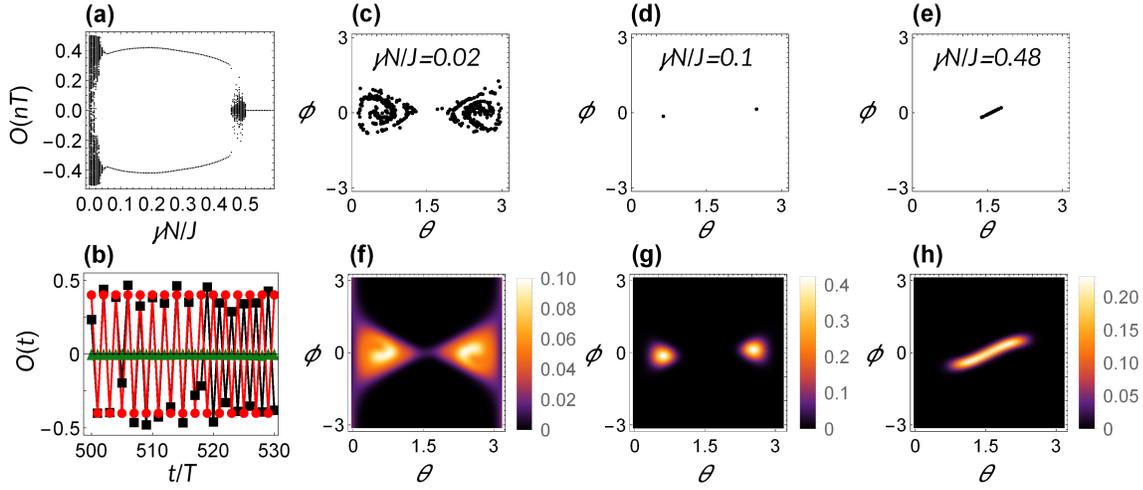}
	\caption{(a) The density imbalance $O(t)$ at time instance  $nT$ ($500<n\leq550$) versus dissipation strength $\gamma N/J$. (b) The evolution of $O(t)$ at stroboscopic time for three typical dissipation strength $\gamma N/J$. The lines with square, dotted and triangle symbols in (b) are for $\gamma N/J=0.02,0.1$ and $0.48$, respectively. (c)-(e) The trajectory of $\{\theta,\phi\}$ at stroboscopic time $nT$ with $500<n\leq 1000$ for the three typical dissipation strength used in (b). The results in (a)-(e) are calculated by mean-field equations and the initial states for the simulations are randomly chosen.  (f)-(h) The projection of the first eigenmode of $\mathcal{V}(T,0)$ to $|\theta, \phi\rangle$ for comparison, color in the figures represent the weight on different $|\theta, \phi\rangle$. The parameters used in (f)-(h) is $N=110$, $\gamma N/J=0.02,0.1$ and $0.48$, respectively. The other parameters used are $J_2=J=1$, $J_1=4J$, $\alpha=0$, $T=2.5\pi/J$, $\xi=\pi/(8J)$ and $UN/J=-4$. }\label{mean}
\end{figure*}

We numerically integrate Eq.(\ref{angle}) to obtain the evolution of $\theta(t)$ and $\phi(t)$. For the case of symmetric double well $\alpha=0$, the values of $O(t)=\cos(\theta)/2$ at $t=nT$ with integer $n$  ($500<n\leq550$) are plotted in Fig.\ref{mean}(a). Similar with the master equation approach, with moderate dissipation, there are stable period doubling limit circles as shown in Fig.\ref{mean}(a). Fig.\ref{mean}(b) shows the evolution of $O(t)$ with three different dissipation strength, line with dotted symbols show the period doubling evolution with moderate dissipation $\gamma N/J=0.1$. The lines with square and triangle symbols in Fig.\ref{mean}(b) show the system in the chaotic regimes with weak and strong dissipation $\gamma N/J=0.02$ and $0.48$, respectively. Fig.\ref{mean}(c)-(e) show the trajectory of $\{\theta,\phi\}$ at stroboscopic time $nT$ ($500<n\leq1000$) for these three different dissipation strength. Fig.\ref{mean}(f)-(h) are their quantum version by projecting the first eigenmode of propagator $\mathcal{V}(T,0)$ to coherent states $|\theta,\phi \rangle=[\cos(\theta/2)\hat{b}_1^{\dagger}+e^{i \phi}\sin(\theta/2)\hat{b}_2^{\dagger}]^N|vac\rangle/\sqrt{N}$ \cite{weiss2008}. The quantum version show similar texture with the mean-field one, except the broadening of texture due to the quantum fluctuations.

\begin{figure}
	\includegraphics[scale=0.6]{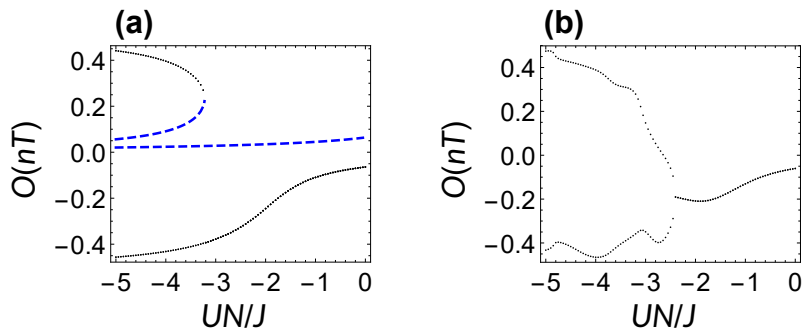}
	\caption{The density imbalance $O(t)$ at time instance $nT$ ($500<n\leq550$) versus the interaction strength $UN/J$. (a) The undriven system, tunneling strength $J(t)=J=1$ is time-independent. (b) The periodic driving system. The other parameters used are the same as in  Fig.\ref{broken} and the initial states are chosen randomly. The dashed blue lines in panel (a) are the unstable real solutions of $\partial_t S_i=0$, $i=x,y,z$. The results in (a) and (b) are obtained by solving the mean-field equations.}\label{brd}
\end{figure}

When the symmetry-breaking points  bias $\alpha\neq0$, the mean-field equation also gives consistent results with the original master equation Eq.(\ref{mastereq}). As shown in Fig.\ref{brd} (a), the undriven system has two stable steady state solution when the interaction strength $UN/J<-3.25$.  Fig.\ref{brd} (b) shows that the period doubling of the driving system appears when $UN/J<-2.4$.

The emergence of DTC in our system without meta-stable state can be understood as follows: Choose the steady state of the undriven system $\rho_1$  as the initial state, and suppose the driving procedure over one period $T$ can be divided into two stages, say  $\mathcal{V}_1$(from $t=0$ to $t=\xi$ in the first $T$, and from $t=T$ to $t=T+\xi$ in the second $T$, and so on) and $\mathcal{V}_2$ (from $t=\xi$ to $t=T$ in the first $T$, and from $t=T+\xi$ to $t=2T$ in the second $T$...). $\mathcal{V}_1$   map $\rho_1$ to a state $\rho_2=\mathcal{V}_1[\rho_1]$ that is not too close to $\rho_1$, and then map $\rho_2$ close to $\rho_1$ again, i.e. $|\rho_2-\rho_1|$ is as large as possible and $|\mathcal{V}_1[\rho_2]-\rho_1|$ is as small as possible, here $|(...)|$ denotes the trace norm of $(...)$, i.e. $|\sigma|=Tr[\sqrt{\sigma^\dagger \sigma}]$. In the second stage,  $\mathcal{V}_2$ is generated by the Lindbladian of the undriven system, for example, $\mathcal{V}_2=\exp[\mathcal{L(\xi)}(T-\xi)]$ in our case. $\mathcal{V}_2$ should map $\rho_2$ not far away from $\rho_2$, i.e. $\rho_3=\mathcal{V}_2[\rho_2]$ and $|\rho_3-\rho_2|$ is close to zero. Then after the   first stage of the second driving period $T$, we have $\rho_4=\mathcal{V}_1\mathcal{V}_2\mathcal{V}_1[\rho_1]=\mathcal{V}_1[\rho_3]$, and  $|\rho_4-\rho_1|=|\rho_4-\mathcal{V}_1[\rho_2]+\mathcal{V}_1[\rho_2]-\rho_1|\leq|\mathcal{V}_1[\rho_3-\rho_2]|+|\mathcal{V}_1[\rho_2]-\rho_1|\approx0$ by the aforementioned spirit of construction. With all mentioned, in the first three stages in the two driving period $2T$, the system  behaves like being driven by  an effective weak drive that steers  the steady state $\rho_1$ slightly from the equilibrium. At the final stage of the first two driving period $2T$, the system was driven from  the state $\rho_4$ to the equilibrium state $\rho_1$ again. Note that $\rho_1$ is the steady state of the system.

When the effective weak drive $\mathcal{V}_1\mathcal{V}_2\mathcal{V}_1$ and the final stage dissipation process $\mathcal{V}_2$  balances well and notice that   $\mathcal{V}_2$ is not required to have   meta-stable state, we can expect that an observable of the system would synchronize with the drive at period $2T$ finally.

\begin{figure}
	\includegraphics[scale=0.4]{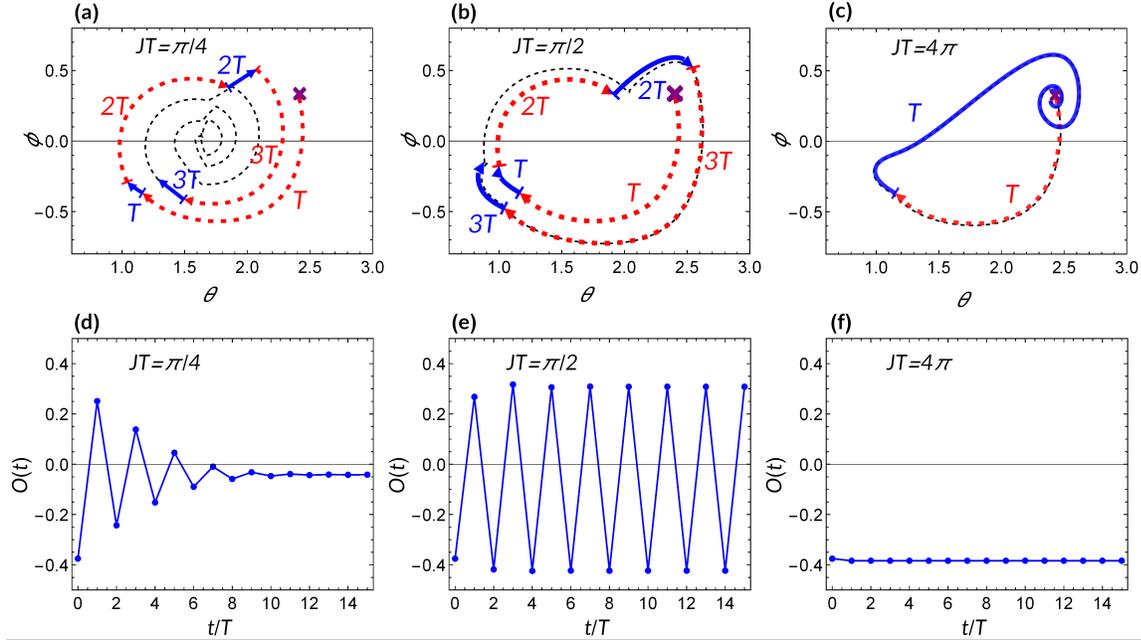}
	\caption{(a)-(c) The trajectory (thin black dash lines) of the quantum system in space $\{\theta,\phi\}$. The result is obtained  by solving the mean-field equation. The dimensionless interaction strength is $UN/J=-3$, so there is no meta-stable state  in the system without drive.
       The purple crosses in (a)-(c) indicate the initial states of the evolution and they are also the equilibrium state of the undriven system.  The thick dash arrow lines indicate the flip process within one driving period when $t<\xi$, while the thick solid lines indicate the dissipation process when  $\xi<t<T$.  Three different driving periods $JT=\pi/4,\pi/2$ and $4\pi$ are chosen, corresponding respectively  to (a)-(c). The labels near the lines indicate the driving periods $nT$. (d)-(f) are the   corresponding density imbalance  $O(t)$ of sub-panel (a)-(c) at integer multiple of period $T$. The other parameters are the same as in Fig.3 (b).}\label{diT}
\end{figure}

Indeed, the aforementioned scheme can be realized theoretically in our system, as  Fig. \ref{diT} (a)-(c) show. In Fig. \ref{diT} (a)-(c), the parameters are chosen such that there is no meta-stable state for the undriven system. The purple crosses in Fig. \ref{diT} (a)-(c) indicate the initial state of the system in the $\{\theta, \phi\}$ parameter space and it is also  the steady state of undriven system. The lines with arrow denote the first (dash line) and the second stage (solid line) of the driving procedure in one period $T$. The number $nT$ ($n=1,2,3$) near the lines denote the $n$-th driving period. In Fig. \ref{diT} (a),  $\mathcal{V}_1$ maps the initial state far away from its initial, the effective driving $\mathcal{V}_1\mathcal{V}_2\mathcal{V}_1$ and dissipation $\mathcal{V}_2$ can not arrive at a  balance, the state of system moves far away from the initial equilibrium position after every two driving period and finally stabilize at a position away from the steady state of the undriven system. In this case, we can not have a DTC. In Fig. \ref{diT} (b), the effective driving and dissipation is properly  balanced, after a transient time, a rigid period-$2T$ oscillation is formed. The state of system returns to a position that is almost  the steady state of the undriven system at $2nT$ times. In Fig. \ref{diT} (c), the drive at the first stage maps the initial state far away from the initial one, but the duration of the second stage of drive is too long, so  the dissipation dominates and the system returns to an  equilibrium state near the steady state of the undriven system, leading to an oscillation with period-$T$. Fig. \ref{diT} (d)-(f) are the dynamics of  $O(t)$   corresponding to   Fig. \ref{diT} (a)-(c) at integer multiple time of period $T$. By this study, we also find that relatively strong interaction is favorable for the required balance between   $\mathcal{V}_1\mathcal{V}_2\mathcal{V}_1$ and  $\mathcal{V}_2$.

\begin{figure}
	\includegraphics[scale=0.55]{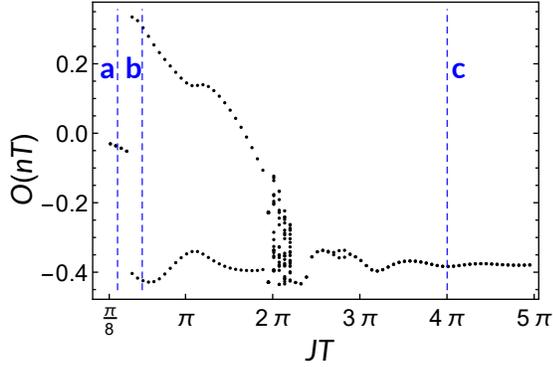}
	\caption{The density imbalance $O(t)$ at time instance $nT$ ($500<n\leq550$) versus the driving period $JT$. The dimensionless interaction strength $UN/J=-3$, the other parameters used are the same as Fig.\ref{broken} (b). and the initial states are chosen randomly. The blue dash lines with labels a,b and c indicate the three different driving periods used in Fig. \ref{diT} (a)-(c), respectively. The result is obtained by solving the mean-field equations.}\label{pT}
\end{figure}

Fig. \ref{pT} shows the values of $O(t)$ at time $nT$ ($500<n\leq550$) with different (dimensionless) driving periods $JT$, the other parameters in Fig. \ref{pT} are chosen as  the same as in Fig. \ref{diT}. The three drives used in Fig. \ref{diT} are indicated by blue dash lines(labeled by a,b,and c) in Fig. \ref{pT}. We can find that with the drive "a", $O(t)$ can take only value at $nT$($500<n\leq550$), indicating that there is no DTC. Whereas  with the drive "b", two values can be taken for  $O(t)$, suggesting that the DTC is of $2T$. The feature of the system with drive "c" is the same as "a". From Fig. \ref{pT}, we observe that except the three typical drives indicated by dash lines (labeled by a, b and c), a wide range of drives can be found that possess the same feature as discussed above. Besides, there is a crossover region between $JT=2\pi$ and $JT=3\pi$(around $2\pi$ in the figure), in this region, the evolution of  $O(t)$ could take many values, this does not mean that there is a DTC  of $nT$, as the values $O(t)$ might   increases if we prolong the observation time(for example the observation time is $nT$ with $500<n\leq 1000$).

\subsection{Mean-field analysis for $2n$ sites model}
\begin{figure}
	\includegraphics[scale=0.65]{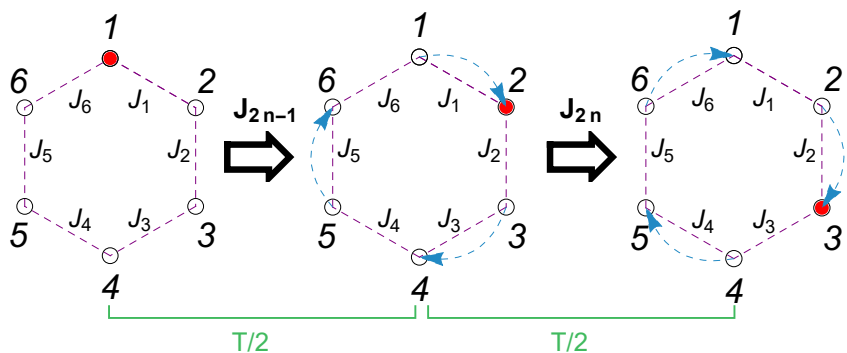}
	\caption{Illustration  of a ring lattice of sites $2n=6$ with a driving lasting over one period $T$. The red dot in the left panel represents the initial state of BECs. The full driving process is formed by two flips marked with blue dashed arrows.}\label{six}
\end{figure}

For the ring lattice with $2n$ sites, the modulation of tunneling are
\begin{equation}
J_o(t)=\left\{
\begin{array}{cc}
J & 0\leq t<T/2-\xi \\
J_f & T/2- \xi\leq t<T/2 \\
J & T/2\leq t<T \\
\end{array}
\right. ,
\end{equation}
and
\begin{equation}
J_e(t)=\left\{
\begin{array}{cc}
J & 0\leq t<T-\xi \\
J_f & T-\xi\leq t<T \\
\end{array}
\right. .
\end{equation}
Fig.\ref{six} is the pictorial representation of the ring lattice of sites $2n=6$. Here, we focus the situation $2n=6$ as an example, but the similar results can be obtained for any $M=2n$.

Intuitively, the two sites model can be regard as the basic building block of the more complex large rings. As shown in Fig.\ref{six}, the condensate is firstly loaded on site $1$. In the first half period, similar with the two sites model, we can flip the condensate from site $1$ to site $2$, then wait some time to trap the atoms in site $2$. In the second half period, we do exactly the same thing but flip the condensate from site $2$ to site $3$. Repeat such procedure, we might observe the clockwise circulation of the particles with period $3T$. If initially load the atoms on even site, the circulation can be anti-clockwise.

To probe such clockwise circulation, we define generalized imbalance
\begin{equation}
O(t)=\sum_{j=0}^{n-1}e^{i2\pi j/n}n_{2j+1}(t).
\end{equation}
Here, $n_{2j+1}$ is the expectation value of particle density at site $2j+1$. The Fourier transform of $O(t)$ is defined by
\begin{equation}
\tilde{O}(\omega)=\lim_{W\rightarrow \infty} \frac{1}{W}\sum_{t=0}^{WT}O(t)e^{-i\omega t}.
\end{equation}
If the atoms rotate clockwise, we have  $O(kT)\sim e^{i2\pi k/n}$, the phase $2\pi k/n$ increase linearly as function of time, and the Fourier spectrum $\tilde{O}(\omega)$ has a sharp peak  at $\omega/n$, i.e. period $nT$ time crystal. For the case $2n=6$, we have $n=3$ and a $3T$ period time crystal. The particle density $n_{2j+1}(t)=|b_{2j+1}(t)|^2$ are calculated by the mean filed equation,
\begin{equation}
	\eqalign{
\partial_t b_l=&-i[\alpha_l b_l+Un_l b_l-(J_l b_{l+1}+J_{l-1}b_{l-1})] \cr
&+2\gamma_{l-1}(n_{l-1}+b_{l-1}^*b_l)(b_{l-1}-b_l) \cr
&-2\gamma_{l}(n_{l+1}+b_{l+1}^*b_l)(b_l-b_{l+1}).}
\end{equation}
This equation is obtained by $b_l(t) \equiv Tr[\hat{b}_l \hat{\rho}(t)]$ and the assumption $Tr[\hat{b}_k^\dagger \hat{b}_l \hat{\rho}(t)]\approx Tr[\hat{b}_k^\dagger\hat{\rho}(t)]Tr[\hat{b}_l \hat{\rho}(t)]$. The mean field approximation is done in the following way. First,  separate the field operator as $\hat{b}_i=b_i+\delta \hat{b}_i$, where $b_i$ is the average value of the field and $\delta \hat{b}_i$ is the corresponding quantum fluctuation. Such approximation is valid when the (second order or high) quantum correlation  is negligible compared to the average value of the fields, i.e. $\langle \delta \hat{b}_i \delta \hat{b}_j \rangle\ll b_i b_j$ (note that $\langle \delta \hat{b}_i \rangle =\langle \delta \hat{b}_j \rangle=0$), this happens when the coupling $J$  between the two sites is small. For two-site model, the mean field approximation shows agreement with the exact one. For large number of lattice sites, exact quantum mechanical treatment  is difficult  due to the huge Hilbert space.  The validity of the mean field approximation with large number of lattice sites is still an open question, but in a recent experimental work that studies driven-dissipative quantum phase transition in a one dimensional Bose-Hubbard chain \cite{fitzpatrick2016},  showed the agreement between the observation and the mean field prediction. Early theoretical work \cite{weiss2008} found that the difference between the non-mean-field results  and the  mean-field one can be reduced by introducing decoherence. This is exactly the scope studied in this work.

\begin{figure}
	\includegraphics[scale=0.5]{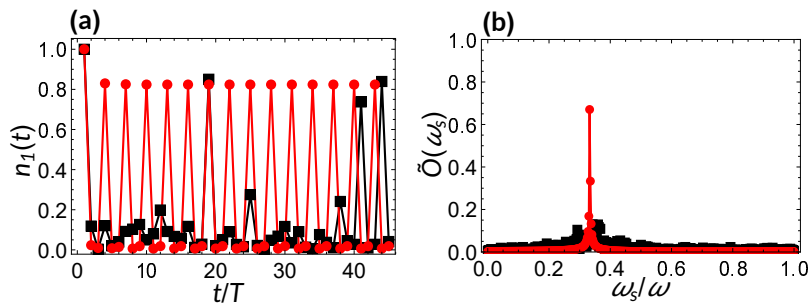}
	\caption{(a) The evolution of the density at the first site in the system of six sites. The parameters used in the calculations are $J=1$, $J_f/J=5$, $UN/J=-5.5$, $T=4\pi/J$ and $\alpha=0.1J$. The line with dotted symbols for $\gamma N/J=0.2$ and line with square symbols for $\gamma N/J=0.1$. (b) The corresponding Fourier spectrum of generalized imbalance, dot and square are for $\gamma N/J=0.2$ and $0.1$. The initial states used in the calculations are $n_1/N=1$ and $n_{l\neq1}/N=0$. }\label{td}
\end{figure}

\begin{figure}
	\includegraphics[scale=0.56]{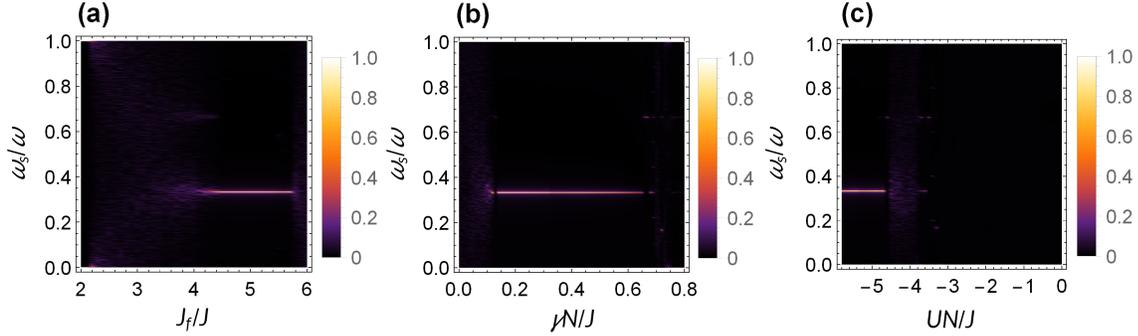}
	\caption{The Fourier spectrum of the generalized imbalance versus various parameters for model with six sites. (a) The Fourier spectrum for $UN/J=-5.5$, $\gamma N/J=0.2$ as a function of  $J_f/J$. (b) The Fourier spectrum for $UN/J=-5.5$, $J_f/J=5$ versus  $\gamma N/J$. (c) The Fourier spectrum for $\gamma N/J=0.2$, $J_f/J=5$ varies with $UN/J$. The other parameters used in the calculations are $J=1$, $T=4\pi/J$ and $\alpha=0.1J$. The initial states used in the calculations are $n_1/N=1$ and $n_{l\neq1}/N=0$.}\label{three}
\end{figure}

In Fig.\ref{td} (a), we plot the evolution of $n_1(t)$ for six sites model. When $\gamma N/J=0.2$, the evolution of $n_1(t)$ show rigid oscillation with period $3T$. For $\gamma N/J=0.1$, the system is in the normal phase, and the evolution $n_1(t)$ is irregular(not periodic). Fig.\ref{td} (b) shows the corresponding Fourier spectrum, when system is in the time crystal phase, there is a sharp peak  at $\omega_s=\omega/3$. In Fig.\ref{three}, we show the Fourier spectrum of generalized imbalance versus various parameters for a system of six sites. It is clear  that the rigid sub-harmonic response with reduced frequency $\omega/3$ is robust to the perturbation in the system parameters. We also find that the result of six sites model is similar to that of two sites model, the time crystal phase emerges when the interaction is strong   and the dissipation  takes a moderate value.

\section{Conclusion and discussions}\label{cd}

To conclude, we study a Bose-Hubbard model with dissipation and periodically tunneling. In the simplest case of only two sites, we study in detail the dynamics by means of Floquet-Lindblad formalism. When the two-site system is in the self-trapping regime, we found that the system can exhibit sub-harmonic dynamical responses to the periodic
drive, forming a DTC.  An periodic drive can also turn the system to a DTC even if there is no self-traping in the system, this is different from the result in the earlier study that requires   meta-stable states in
the undriven system. Furthermore, we have shown that period-$n$ ($n > 2$)
DTC can be realized by using two sites model as basic building blocks to construct large rings. The present results  might find applications into engineering exotic phases in driven open quantum systems.

\section*{ACKNOWLEDGMENTS}
This work is supported by National Natural Science Foundation of
China (NSFC) under Grants No. 11775048, No. 11534002 and No.11947405.


\Bibliography{0}

\bibitem{wilczek2012} Wilczek F 2012 Phys. Rev. Lett. \textbf{109} 160401
\bibitem{li2012} Li Tongcang, Gong Zhe-Xuan, Yin Zhang-Qi, Quan H T, Yin Xiaobo, Zhang Peng, Duan L M, and Zhang Xiang 2012 Phys. Rev. Lett. \textbf{109} 163001
\bibitem{watanabe2015} Watanabe H and Oshikawa M 2015 Phys. Rev. Lett. \textbf{114} 251603
\bibitem{syrwid2017} Syrwid A, Zakrzewski J, and Sacha K 2017 Phys. Rev. Lett. \textbf{119} 250602
\bibitem{else2016} Else D V, Bauer B, and Nayak C 2016 Phys. Rev. Lett. \textbf{117} 090402
\bibitem{yao2017} Yao N Y, Potter A C, Potirniche I D, and Vishwanath A 2017 Phys. Rev. Lett. \textbf{118} 030401
\bibitem{huang2018} Huang Biao, Wu Ying-Hai, and Vincent Liu W 2018 Phys. Rev. Lett. \textbf{120} 110603
\bibitem{giergiel2019} Giergiel K, Kuroś A, and Sacha K 2019 Phys. Rev. B \textbf{99} 220303(R)
\bibitem{lazarides2017} Lazarides A and Moessner R 2017 Phys. Rev. B \textbf{95} 195135
\bibitem{zhang2017} Zhang J, Hess  P W, Kyprianidis A, Becker P, Lee A, Smith J, Pagano G, Potirniche I D, Potter A C, Vishwanath A, et al. 2017 Nature (London) \textbf{543} 217
\bibitem{choi2017} Choi S, Choi J, Landig R, Kucsko G, Zhou H, Isoya J, Jelezko F, Onoda S, Sumiya H, Khemani V, et al. 2017 Nature (London) \textbf{543} 221
\bibitem{gong2018} Gong Z, Hamazaki R, and Ueda M, 2018 Phys. Rev. Lett. \textbf{120} 040404
\bibitem{gambetta2019}  Gambetta F M, Carollo F, Marcuzzi M, Garrahan J P, and Lesanovsky I, 2019 Phys. Rev. Lett. \textbf{122} 015701
\bibitem{baumann2011} Baumann K, Mottl R, Brennecke F, and Esslinger T, 2011 Phys. Rev. Lett. \textbf{107} 140402
\bibitem{smerzi1997} Smerzi A, Fantoni S, Giovanazzi S, and  Shenoy S R 1997 Phys. Rev. Lett. \textbf{79} 4950
\bibitem{wang2007} Wang W, Fu L B, and Yi X X, 2007 Phys. Rev. A \textbf{75} 045601
\bibitem{diehl2008} Diehl S,Micheli A, Kantian A, Kraus B,  B\"{u}chler H P, and  Zoller P, 2008 Nat. Phys. \textbf{4} 878
\bibitem{bloch2008} Bloch I, Dalibard J, and Zwerger W, 2008 Rev. Mod. Phys. \textbf{80} 885
\bibitem{giergiel2018} Giergiel K, Kosior A, Hannaford P, and Sacha K 2018 Phys. Rev. A \textbf{98} 013613
\bibitem{blanes2009} Blanes S,  Casas F,  Oteo J A, et al., 2009 Phys. Rep. \textbf{470} 151-238
\bibitem{dai2016} Dai C M, Shi Z C, and Yi X X, 2016 Phys. Rev. A \textbf{93} 032121
\bibitem{vega2017} de Vega I and Alonso D, 2017 Rev. Mod. Phys. \textbf{89} 015001
\bibitem{hartmann2017} Hartmann M,  Poletti D,  Ivanchenko M, Denisov S,  and Hänggi P, 2017 New J. Phys. \textbf{19} 083011
\bibitem{albert2014} Albert V V and Jiang L, 2014 Phys. Rev. A \textbf{89} 022118
\bibitem{arnol1992} Arnol'd  V I, 1992 Ordinary Differential Equations, Springer-Verlag Berlin Heidelberg
\bibitem{autti2018} Autti S, Eltsov V B, and Volovik G E, 2018 Phys. Rev. Lett.120 215301
\bibitem{cui2010} Cui B, Wang L C, and Yi X X, 2010 Phys. Rev. A 82 062105
\bibitem{watanabe2012} Watanabe G and M\"{a}kel\"{a} H, 2012 Phys. Rev. A \textbf{85} 053624
\bibitem{weiss2008} Weiss C and Teichmann N, 2008 Phys. Rev. Lett. \textbf{100} 140408
\bibitem{fitzpatrick2016} Fitzpatrick M, Sundaresan N M, Li A C Y, Koch J, and Houck A A 2016 Phys. Rev. X \textbf{7} 011016
\endbib
\end{document}